\documentclass[
 longbibliography,
 twocolumn,
 superscriptaddress,
 amsmath,
 amssymb,
 amsthm,
 amsfonts,
 floatfix
]{revtex4-2}

\usepackage[colorlinks=true,urlcolor=blue,citecolor=blue,linkcolor=blue]{hyperref} 
\usepackage{graphicx}
\usepackage{dcolumn}
\usepackage{kotex}
\usepackage{natbib}
\usepackage{color}
\usepackage{amsfonts}
\usepackage{hyperref}
\usepackage{bm}
\usepackage{pdfpages}
\usepackage{pgffor}
\makeatletter
\AtBeginDocument{\let\LS@rot\@undefined}
\makeatother

\newcommand{\RNum}[1]{\uppercase\expandafter{\romannumeral #1\relax}}

\begin{document}

\title{Attaining entropy production and dissipation maps \\from Brownian movies via neural networks}

\author{Youngkyoung Bae}
 \affiliation{Department of Physics, Korea Advanced Institute of Science and Technology, Daejeon 34141, Korea}
 
 \author{Dong-Kyum Kim}
 \affiliation{Department of Physics, Korea Advanced Institute of Science and Technology, Daejeon 34141, Korea}

\author{Hawoong Jeong}
\email{hjeong@kaist.edu}
\affiliation{Department of Physics, Korea Advanced Institute of Science and Technology, Daejeon 34141, Korea}
\affiliation{Center for Complex Systems, Korea Advanced Institute of Science and Technology, Daejeon 34141, Korea}

\date{\today}

\begin{abstract}

Quantifying entropy production (EP) is essential to understand stochastic systems at mesoscopic scales, such as living organisms or biological assemblies. However, without tracking the relevant variables, it is challenging to figure out where and to what extent EP occurs from recorded time-series image data from experiments. Here, applying a convolutional neural network (CNN), a powerful tool for image processing, we develop an estimation method for EP through an unsupervised learning algorithm that calculates only from movies. Together with an attention map of the CNN’s last layer, our method can not only quantify stochastic EP but also produce the spatiotemporal pattern of the EP (dissipation map). We show that our method accurately measures the EP and creates a dissipation map in two nonequilibrium systems, the bead-spring model and a network of elastic filaments. We further confirm high performance even with noisy, low spatial resolution data, and partially observed situations. Our method will provide a practical way to obtain dissipation maps and ultimately contribute to uncovering the nonequilibrium nature of complex systems.

\end{abstract}

\maketitle

\section{Introduction}

From large-scale active matter to living organisms on the nanoscale, most systems in nature are not in equilibrium but rather in the nonequilibrium state~\cite{marchetti2013hydrodynamics, fang2019nonequilibrium}.
As the development of experimental techniques has led us to more closely observe mesoscopic systems, in which thermal fluctuations dominate their dynamics, how we can describe and measure thermodynamic quantities in such stochastic systems has been actively studied. 
These systems, such as beating flagella or cilia~\cite{vilfan2006hydrodynamic, battle2016broken, saggiorato2017human}, molecular motors like kinesin walking along microtubules~\cite{yildiz2004kinesin, yildiz2008intramolecular, ariga2018nonequilibrium}, motile bacteria~\cite{cates2012diffusive, mathijssen2019oscillatory}, and cytoskeletal networks~\cite{mizuno2007nonequilibrium, seara2018entropy}, maintain their nonequilibrium activity by exchanging energy and information with their surrounding environment or consuming chemical fuels, e.g., ATP hydrolysis, and then converting it into mechanical work.
One of the principal ways to characterize or identify the nonequilibrium nature of systems is by measuring the energy dissipated into the environment through these processes~\cite{harada2005equality, toyabe2010nonequilibrium, lander2012noninvasive, ariga2018nonequilibrium}.
This dissipated energy, or dissipation, can be represented by entropy production (EP), a measure of the system's irreversibility.
As the connection between irreversibility and dissipation on the trajectory level has been established, EP has become an essential quantity in comprehending the stochastic dynamics of systems~\cite{kawai2007dissipation, parrondo2009entropy, seifert2012stochastic, gnesotto2018broken, dabelow2019irreversibility}.

With the growing interest in EP, numerous studies have been conducted to infer the EP rate from recorded data of the system.
Obtaining the empirical probability current in the coarse-grained space of tracked variables is a conventional method for quantifying nonequilibrium activity from time-series images or movies~\cite{battle2016broken, gladrow2016broken, gnesotto2018broken}.
Some methods based on the thermodynamic uncertainty relation~\cite{barato2015thermodynamic, Li2019quantifying, manikandan2019inferring, van2020entropy, otsubo2020estimating, otsubo2021estimating, kamijima2021higherorder} as well as the neural estimator for entropy production (NEEP)~\cite{kim2020learning, otsubo2021estimating} have been recently proposed for continuous trackable variables and have successfully inferred the EP rate from the trajectories.
However, these previous attempts have a limitation in that they are not applicable to systems in which the configurational state variables cannot be tracked due to a poor spatiotemporal resolution of the movies or system fragility in attaching heavy tracer particles~\cite{manzo2015review}.
Moreover, even though the advent of fluorescence microscopy techniques has enabled us to observe the dynamic processes of various living systems directly~\cite{brangwynne2007bending, waters2009accuracy, manzo2015review, seara2018entropy, liu2020quantifying}, finding which variables are relevant to their dynamics or which regions reveal the nonequilibrium activity yet remain as interesting open questions in real complex biological systems.
A stochastic force inference method has been applied to time-series images using a dimensionality reduction approach~\cite{frishman2020learning, gnesotto2020learning} to address these issues, but this is hard to apply in high-dimensional systems and is unable to capture how much EP occurred in which area of the image.

In this work, we propose the convolutional neural estimator for entropy production (CNEEP) for estimating stochastic EP from time-series images, called Brownian movies, based on the unsupervised learning algorithm recently developed by our group~\cite{kim2020learning}.
Specifically, without detailed information about the system, such as drift or diffusion fields and which regions contain relevant variables, the CNEEP can provide a {\it dissipation map}, which embodies where and how much EP occurs at each transition in the images.
The dissipation map can bring out the contribution that each region has to the system's irreversibility, helping to grasp the heterogeneous spatial pattern of nonequilibrium phenomena~\cite{nardini2017entropy, guo2021play}.
Our approach is the first study to our knowledge that directly produces a dissipation map from movies without any extraction procedure of relevant components.
To validate our method, we apply the CNEEP to the bead-spring model as well as biopolymer networks for a high-dimensional complex case.
Finally, we show that our method is also applicable to incomplete scenarios, namely when some measurement noise or a low spatial resolution of a microscope hinders us from detecting the exact information of the system, or when we can only observe a partial region of a whole system.

\section{Description of our approach}
\label{sec:2}

\begin{figure}[!b]
    \includegraphics[width=\linewidth]{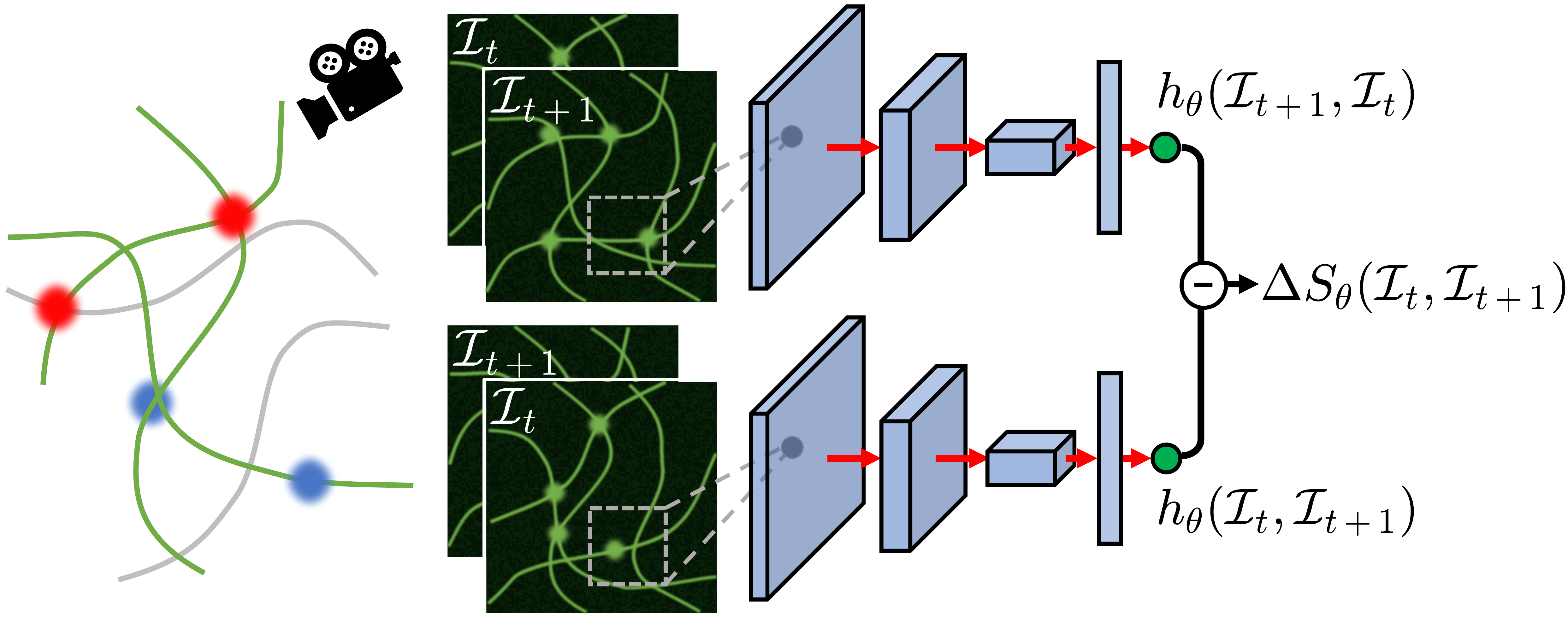}
    \vskip -0.1in
    \caption{ Schematic of the convolutional neural estimator for entropy production (CNEEP). Feeding an image set of a forward transition ($\mathcal{I}_t$, $\mathcal{I}_{t+1}$) and its reversed set ($\mathcal{I}_{t+1}$, $\mathcal{I}_{t}$) to a neural network, $h_{\theta}\left(\mathcal{I}_t, \mathcal{I}_{t+1} \right)$ and $h_{\theta}\left(\mathcal{I}_{t+1}, \mathcal{I}_{t} \right)$ are obtained as their respective outputs. The final output of CNEEP is calculated by $\Delta S_\theta \left(\mathcal{I}_t, \mathcal{I}_{t+1} \right) \equiv h_{\theta}\left(\mathcal{I}_t, \mathcal{I}_{t+1} \right) - h_{\theta}\left(\mathcal{I}_{t+1}, \mathcal{I}_{t} \right)$.
    }\label{fig1}
\end{figure}

We describe our method to estimate the stochastic EP and create a dissipation map from time-series image data.
Let us denote the recorded time-series images at each time interval $\Delta \tau$ from a steady-state system as $\{ \mathcal{I}_0, \mathcal{I}_1, \dots, \mathcal{I}_{L-1} \}$, where $L$ is the number of steps. The image at time-step $t$ denoted by $\mathcal{I}_t$ is a transformed representation of the configurational state vector $\bm{z}(\tau)$ at time $\tau \equiv t \Delta \tau$ into a $(W, H)$ array. 
Each element of the array (a pixel intensity) has an integer value ranging from $0$ to $255$, like the form of an actual image.
If we assume that the underlying system is Markovian and $\mathcal{I}_t$ contains the full information of $\bm{z}(\tau)$, in other words the transformation from $\bm{z}(\tau)$ to $\mathcal{I}_t$ is an invertible mapping, then the stochastic EP $\Delta S$ between time-step $t$ and $t+1$ can be determined by $\mathcal{I}_t$ and $\mathcal{I}_{t+1}$. From this fact, we can build the output of our estimator as
\begin{align} \label{eq:estimator}
    \Delta S_{\theta}\left( \mathcal{I}_t, \mathcal{I}_{t+1} \right) \equiv h_{\theta}\left(\mathcal{I}_t, \mathcal{I}_{t+1} \right) - h_{\theta} \left( \mathcal{I}_{t+1}, \mathcal{I}_{t} \right).
\end{align}
Here, we employ a convolutional neural network (CNN) as $h_\theta$, which is a deep neural network based on convolution filters that has been proven to be powerful in the processing of image datasets~\cite{lecun1998gradient, goodfellow2016deep}, where $\theta$ is the trainable parameters of the CNN.
As shown in Fig.~\ref{fig1}, $h_{\theta}\left(\mathcal{I}_t, \mathcal{I}_{t+1}\right)$ is the output of the CNN for the input $\left(\mathcal{I}_t, \mathcal{I}_{t+1}\right)$. 
Note that $\Delta S_\theta \left(\mathcal{I}_t, \mathcal{I}_{t+1}\right) = -\Delta S_\theta \left( \mathcal{I}_{t+1}, \mathcal{I}_{t} \right)$.
To achieve that $\Delta S_\theta \left(\mathcal{I}_t, \mathcal{I}_{t+1}\right)$ becomes $\Delta S\left(\mathcal{I}_t, \mathcal{I}_{t+1}\right)$, we use the objective function $J(\theta)$ given by
\begin{align}
    J(\theta) &=  \langle \Delta S_{\theta} - e^{-\Delta S_{\theta}} 
    \rangle,
    \label{eq:lossftn}
\end{align}
where the angle bracket $\langle \cdot \rangle$ is the ensemble average.
This objective function has been proposed in our previous work~\cite{kim2020learning}, and it can also be derived from the $f$-divergence representation of Kullback--Leibler divergence~\cite{nguyen2010estimating, belghazi2018mutual, otsubo2021estimating, kamijima2021higherorder}.
The $J(\theta)$ is maximized when $\Delta S_\theta$ is given by
\begin{align}
    \Delta S_\theta \left(\mathcal{I}_t, \mathcal{I}_{t+1} \right) = \ln{ \frac{ \mathcal{P}\left(\mathcal{I}_t, \mathcal{I}_{t+1} \right)}{\mathcal{P}\left(\mathcal{I}_{t+1}, \mathcal{I}_{t} \right)}},
    \label{eq:sol}
\end{align}
where $\mathcal{P}\left(\mathcal{I}_t, \mathcal{I}_{t+1} \right)$ is the probability of observing the transition $\mathcal{I}_t \rightarrow \mathcal{I}_{t+1}$. The right-hand-side is the definition of $\Delta S(\mathcal{I}_t, \mathcal{I}_{t+1})$, and thus the output of our estimator converges to $\Delta S \left(\mathcal{I}_t, \mathcal{I}_{t+1}\right)$ as maximizing $J(\theta)$. Please refer to Supplementary Notes 2, 3 for the training details and configurations of the CNN.

We need an additional step to create a dissipation map from Brownian movies. 
Let $A^{f}_c(x, y)$ ($A^{r}_c(x, y)$) denote an element of the $c$-th channel at spatial position $(x, y)$ in an attention map of the CNN's last convolutional layer for the forward (reversed) transition. Defining $h_\theta \left( \mathcal{I}_t, \mathcal{I}_{t+1} \right)$ as a sum of the elementwise product between $A^{f}_c(x, y)$ and the weight of each element of $c$-th channel denoted by $w_c (x, y)$, then $h_\theta \left( \mathcal{I}_t, \mathcal{I}_{t+1} \right)$ can be written as
\begin{equation}
\begin{aligned}
    h_\theta \left( \mathcal{I}_t, \mathcal{I}_{t+1} \right) &= \sum_{x, y} \mathcal{H}^f(x, y),\\ 
    h_\theta \left( \mathcal{I}_{t+1}, \mathcal{I}_{t} \right) &= \sum_{x, y} \mathcal{H}^r(x, y),
    \label{eq:diss_map}
\end{aligned}
\end{equation}
where $\mathcal{H}^f(x, y) \equiv \sum_{c} w_c(x, y) A^{f}_c (x, y)$ and $\mathcal{H}^r(x, y) \equiv \sum_{c} w_c(x, y) A^{r}_c (x, y)$. According to Eq.~\eqref{eq:estimator}, $\Delta S_{\theta}\left(\mathcal{I}_t, \mathcal{I}_{t+1}\right)$ can thus be obtained as
\begin{align}
    \Delta S_{\theta}\left( \mathcal{I}_t, \mathcal{I}_{t+1} \right) &= \sum_{x, y} \mathcal{H}(x, y), \label{eq:diss_map2}
\end{align}
where $\mathcal{H}(x, y) \equiv \mathcal{H}^f(x, y) - \mathcal{H}^r(x, y)$ and $\mathcal{H}(x, y)$ indicates the spatial importance of $\Delta S_{\theta}\left( \mathcal{I}_t, \mathcal{I}_{t+1} \right)$ at $(x, y)$.
Therefore, $\mathcal{H}(x, y)$ can provide both where and how much EP occurs at $(x, y)$ during the transition $\mathcal{I}_t \rightarrow \mathcal{I}_{t+1}$ when $\Delta S_{\theta}\left( \mathcal{I}_t, \mathcal{I}_{t+1} \right)$ equals to $\Delta S\left( \mathcal{I}_t, \mathcal{I}_{t+1} \right)$ by training.
We identify the spatial pattern of EP via CNEEP and show that $\mathcal{H}(x, y)$ indeed reflects the spatial pattern of EP in the following examples.

Before we go any further, one may point out that many systems in real experiments are not Markovian due to some experimental noise, several inaccessible relevant variables, or a low spatiotemporal resolution of a microscope, though the underlying dynamics are Markovian~\cite{esposito2012stochastic, mehl2012role, shiraishi2015fluctuation, polettini2017effective, skinner2020improved}. In such cases, although estimation of the true EP for the whole system is generally unavailable, inferring the EP from partial information is still important as it can give a lower bound of the true EP as well as clues to the underlying system~\cite{gomez2008lower, Roldan2010Estimating, esposito2012stochastic, martinez2019inferring, skinner2020improved}; for instance, identifying the nonequilibrium mechanism behind the switching behavior of the {\it E. coli} flagellar motor~\cite{korobkova2006hidden, Tu2008nonequilibrium}.
We demonstrate how the CNEEP can relieve such issues preventing estimation in Sec.~\ref{sec:5}.

\begin{figure}[!b]
    \includegraphics[width=\linewidth]{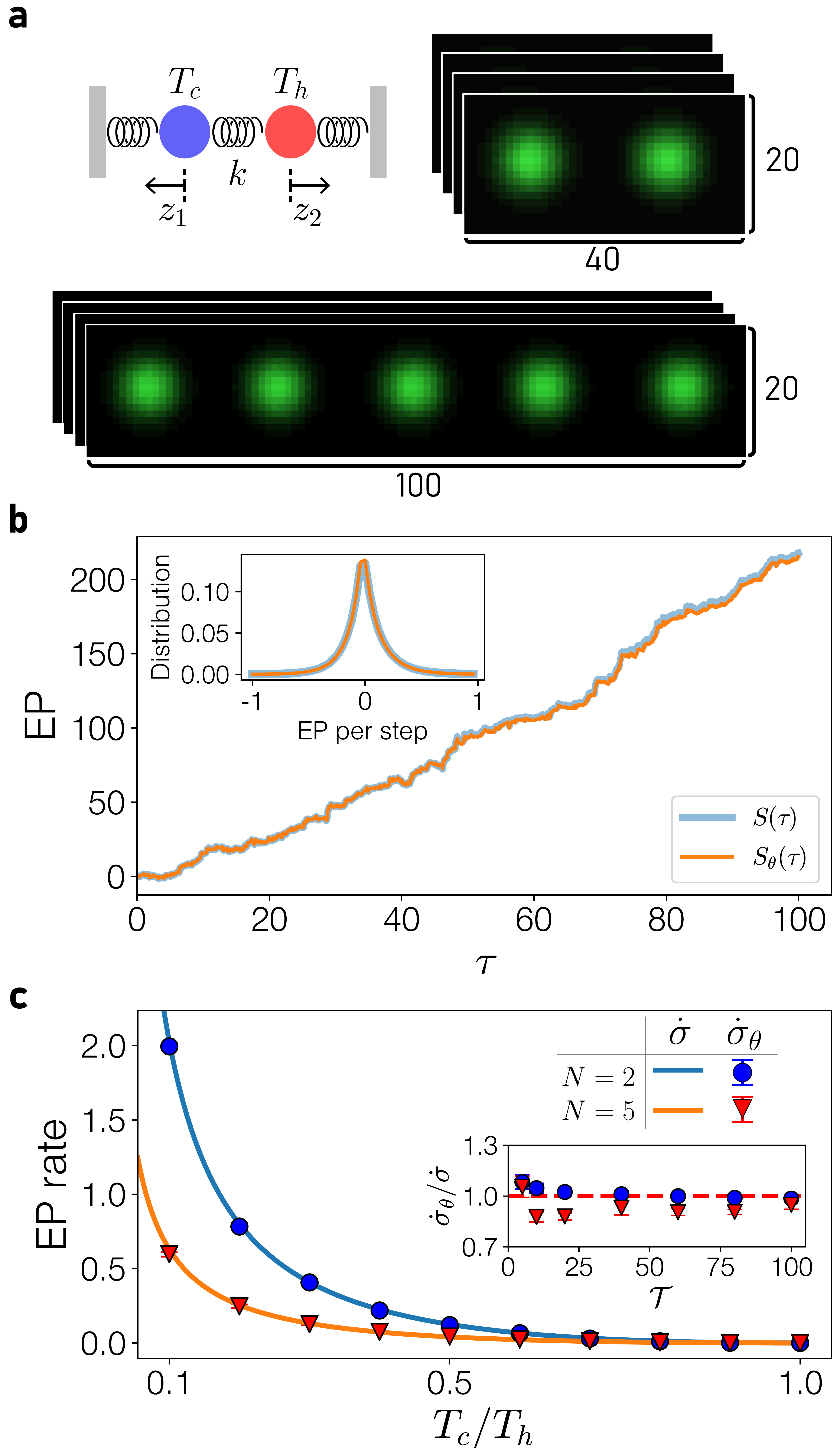}
    \vskip -0.1in
    \caption{{\bf a.} Schematic of a bead-spring model (top left) and generated images of models with $N=2$ (top right) and $N=5$ (bottom). Each image is an integer-valued array with $(20, 20 N)$ pixels.
    {\bf b.} Accumulated EP $S(\tau)$ over time $\tau$ along a single movie for $N=2$, where the orange line is the estimated EP and the blue line is the true EP. The inset represents the distributions of $\Delta S$ and $\Delta S_\theta$ over $100$ movies from the training set.
    {\bf c.} EP rate as a function of $T_c/T_h$ for the $N=2$ and $N=5$ models. The solid lines (symbols) depict the true EP rate $\dot{\sigma}$ (estimated EP rate $\dot{\sigma}_\theta$). The inset shows that $\dot{\sigma}_\theta/\dot{\sigma}$ converges to $1$ with increasing $\mathcal{T}$. Error bars represent the standard deviation of $\dot{\sigma}_\theta$ from five independent estimators.
    }\label{fig2}
\end{figure}
\begin{figure*}[!t]
    \includegraphics[width=\linewidth]{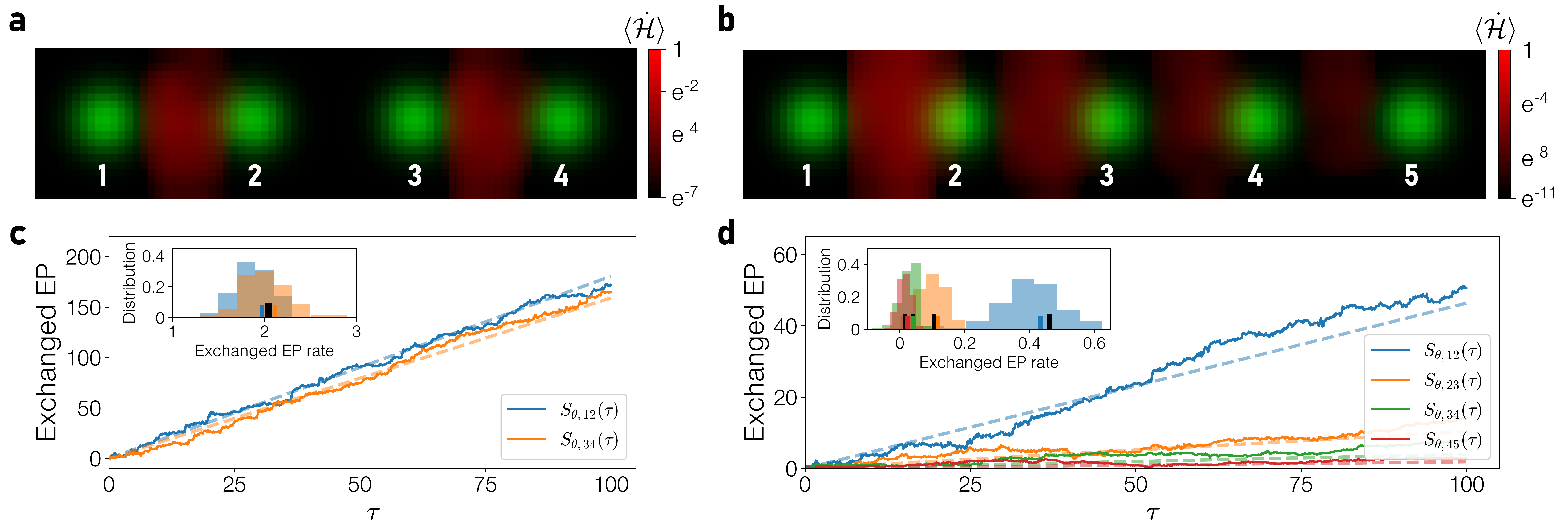}
    \vskip -0.1in
    \caption{
    {\bf a}, {\bf b.} Dissipation map $\langle \dot{\mathcal{H}} \rangle$ (red) overlaid with a snapshot (green) of the bead-spring model with ({\bf a}) $N=4$ and ({\bf b}) $N=5$. In ({\bf a}), there is no connection between beads $2$ and $3$ in the underlying system.
    {\bf c}, {\bf d.} Accumulated exchanged EPs between beads $i$ and $j$ over time $\tau$, denoted by $S_{\theta, ij}(\tau)$, along a single movie.
    The dashed lines indicate the true lines with slope $S_{ij}(\mathcal{T})/\mathcal{T}$, where $\mathcal{T}$ is the recorded time.
    The insets plot the distribution of the exchanged EP rate $\dot{S}_{\theta, ij}$ over $100$ movies. The short solid lines on the bottom of the insets are the estimations of exchanged EP rates $\dot{\sigma}_{\theta, ij}$ [matched colors with $S_{\theta, ij}(t)$] and the true exchanged EP rates $\dot{\sigma}_{ij}$ (black). 
    }
    \label{fig3}
\end{figure*}

\section{Bead-spring model}
\label{sec:3}

As a minimal model to test our approach, we consider the bead-spring model, which consists of $N$ beads connected to the nearest neighbors or boundaries by springs with stiffness $k$ in one dimension. All beads contact different thermal heat baths with friction $\gamma$. The $i$-th bead contacts a bath with temperature $T_i$, which linearly varies from $T_c$ to $T_h$~\cite{chun2015hidden, battle2016broken, Li2019quantifying, gnesotto2020learning}. Then, the system is governed by an overdamped Langevin equation given by
\begin{align}
    \gamma \dot{\bm{z}}(\tau) = \bm{F} \left( \bm{z}(\tau) \right) + \bm{\xi}(\tau),
    \label{eq:Langevin}
\end{align}
where $\bm{z} \equiv (z_1, z_2, \dots, z_N)^T$ is the displacements of the beads, $\bm{F} \left( \bm{z} \right) \equiv \mathsf{A}\bm{z}$ is the drift force with $\mathsf{A}_{ij} = (\delta_{i, j+1} + \delta_{i+1, j} - 2 \delta_{i, j}) k$, and $\bm{\xi}$ is a Gaussian white noise vector satisfying $\langle \xi_i(\tau) \rangle = 0$ and $\langle \xi_i(\tau) \xi_j(\tau') \rangle = 2 \gamma T_i \delta_{i, j} \delta(\tau-\tau')$. 
We set Boltzmann's constant $k_B = 1$.

To make Brownian movies of the bead-spring models, we numerically generated samples of $100$ ($25$) positional trajectories for the training (validation) set with $\Delta \tau = 10^{-2}$ for $N=2$ and $N=5$ in steady-state. The $L$ of each trajectory is $10^4$ (i.e., the recorded time $\mathcal{T} \equiv L\Delta \tau = 100$). The displacement of the beads $\bm{z}(\tau)$ is normalized with a standard deviation $std = 1$ and then transformed into a Gaussian function whose center is located at the corresponding bead's position on an image $\mathcal{I}_t$ as shown in Fig.~\ref{fig2}a; see the details in Methods and the generated movies in Supplementary Movies (SMs) 1, 3.

First, we present the training results for the Brownian movies with $N=2$ at $T_c/T_h = 0.1$ in Fig.~\ref{fig2}b, which exhibits accumulated $\Delta S$ and $\Delta S_\theta$ for a single trajectory, denoted by $S(\tau) \equiv \sum_{t=0}^{\tau/\Delta \tau} \Delta S \left( \mathcal{I}_t, \mathcal{I}_{t+1} \right)$.
As can be seen, $S_\theta(\tau)$ is in line with $S(\tau)$ over time $\tau$, and we also verify that $\Delta S_\theta$ is precisely fitted with $\Delta S$ by $R^2 = 0.9902$ ($R^2 = 0.9896$ on average over all the movies, Supplementary Note 2).
The inset of Fig.~\ref{fig2}b reveals that the distributions of $\Delta S$ and $\Delta S_\theta$ are almost overlapped with each other, supporting that our estimator can exactly provide $\Delta S$ per every transition.

By averaging the slope of $S_\theta (\tau)$ over the Brownian movies, the averaged EP rate $\dot{\sigma}_\theta \equiv \langle S_\theta(\mathcal{T})/\mathcal{T} \rangle$ is obtained and plotted while varying $T_c/T_h$ from $0.1$ to $1$ with $T_h = 10$ for both $N=2$ and $N=5$ (Fig.~\ref{fig2}c). 
We apply five independently trained estimators with different initial parameters to the given range of $T_c/T_h$ and confirm that all the estimators successfully measure $\dot{\sigma}$ with small errors. These small errors suggest that our estimator is robust against altered initial parameters. The inset of Fig.~\ref{fig2}c shows the convergence of $\dot{\sigma}_\theta$ to $\dot{\sigma}$ with increasing $\mathcal{T}$.

To show that our method can create a detailed dissipation map from Brownian movies, we make additional movies that consist of two independent bead-spring models with $N=2$ as shown in Fig.~\ref{fig3}a; in other words, there is no connection between beads $2$ and $3$, while beads $1$ and $2$ (or $3$ and $4$) are connected with an invisible spring---we call this case a bead-spring model with $N=4$. 
Figure~\ref{fig3}a, b show the resulting dissipation map for the $N=4$ and $N=5$ models, respectively, where $\langle \dot{\mathcal{H}}(x, y) \rangle \equiv \langle \mathcal{H}(x, y)/\Delta \tau \rangle$. Note that the spatial sum of $\langle \dot{\mathcal{H}}(x, y) \rangle$ is equal to $\dot{\sigma}_\theta$. For the $N=4$ model, as we expected, $\langle \dot{\mathcal{H}}(x, y) \rangle$ is markedly concentrated between beads $1$ and $2$ ($3$ and $4$), whereas $\langle \dot{\mathcal{H}}(x, y) \rangle$ vanishes between beads $2$ and $3$.
Since this result is in good agreement with the fact that heat flows can exist only between connected beads, it implies that $\langle \dot{\mathcal{H}}(x, y) \rangle$ reflects the actual dissipation map. Please see SMs 2, 3 for $\mathcal{H}(x, y)$ along the $N=4$ and $N=5$ movies.

To examine $\mathcal{H}(x, y)$ quantitatively, we analytically calculate the exchanged EP rate $\dot{\sigma}_{i, i+1}$ occurring between the beads $i$ and $i+1$.
Using formulations based on Eq.~\eqref{eq:Langevin} and a framework for stochastic energetics~\cite{sekimoto2010stochastic, seifert2012stochastic}, $\dot{\sigma}_{i, i+1}$ can be obtained by
\begin{align}
    \dot{\sigma}_{i, i+1} = \langle \dot{Q}_{i, i+1} \rangle \left( \frac{1}{T_i} - \frac{1}{T_{i+1}} \right),
    \label{eq:XEP}
\end{align}
where $\langle \dot{Q}_{i, i+1} \rangle$ is the exchanged heat between beads $i$ and $i+1$~\cite{jarzynski2004classical, berut2016stationary}.
In the linear system, $\langle \dot{Q}_{i, i+1} \rangle$ can be obtained by (Supplementary Note 1)
\begin{align}
    \langle \dot{Q}_{i, i+1} \rangle = \frac{k}{2} \left( \mathsf{A}\mathsf{C} - \mathsf{C}\mathsf{A}^T \right)_{i, i+1},
    \label{eq:XQ}
\end{align}
where the covariance matrix $\mathsf{C} \equiv \langle \bm{x} \bm{x}^T \rangle$.
Here, $\langle \dot{Q}_{i, i+1} \rangle$ is proportional to the anti-symmetric part of $\mathsf{AC}$, which is related to the area enclosing rate or angular momentum~\cite{ghanta2017fluctuation, mura2018nonequilibrium, bae2021inertial}.

We measure the accumulated exchanged EP for both $N=4$ and $N=5$, denoted by $S_{\theta, ij}(\tau)$, by spatially summing $\mathcal{H}(x, y)$ between beads $i$ and $j$ along a single movie.
Because it is difficult to divide $S_{\theta, ij}(\tau)$ into system entropy and exchanged EP, we compare $S_{\theta, ij}(\tau)$ with the line of slope $S_{ij}(\mathcal{T})/\mathcal{T}$ to cancel out the contribution of system entropy.
As shown in Fig.~\ref{fig3}c, d, all the estimations of $S_{\theta, ij}(\tau)$ follow the matched lines along the movie for both models, and especially for the $N=5$ model, the CNEEP well measures the exchanged EPs that occur with varying rates in different areas.
The insets depict the distributions of exchanged EP rates $\dot{S}_{\theta, ij} \equiv S_{\theta, ij}(\mathcal{T})/\mathcal{T}$ over the movies as well as $\dot{\sigma}_{\theta, ij} \equiv \langle \dot{S}_{\theta, ij} \rangle$. As can be seen, all $\dot{\sigma}_{\theta, ij}$ are close to the corresponding $\dot{\sigma}_{ij}$ for both models, which corroborates the ability of our approach to visualize the dissipation of the system.
From these results, therefore, we verify that the CNEEP can provide stochastic EP from movies and create a dissipation map without any knowledge of the interactions between the system components.

\section{Network of elastic filaments}
\label{sec:4}

To illustrate the validity of our method for high-dimensional systems, we apply it to a network of elastic filaments, which has been proposed to study soft biological elastic assemblies driven into a nonequilibrium state by spatially heterogeneous stochasticities~\cite{mura2018nonequilibrium, gradziuk2019scaling, gnesotto2020learning}. An $N^2$ filamentous network comprises $N \times N$ nodes and a number of elastic filaments of stiffness $k$ connecting adjacent nodes or fixed boundaries. 
Because all the nodes and filaments are in a two-dimensional plane, the degree of freedom of the underlying system is $2N^2$. 
It is assumed that the whole system is immersed in a viscous fluid of friction $\gamma$ and temperature $T_c$, while due to enzymatic activity, some nodes can have an additional fluctuation that induces a higher temperature $T_h$ with probability $p$. 
These spatially heterogeneous stochasticities create a heat exchange between connected nodes of different temperatures. 
We suppose that the drift forces are linear in the displacements of nodes $\bm{z}$ to be calculable. 

\begin{figure}[!t]
    \includegraphics[width=\linewidth]{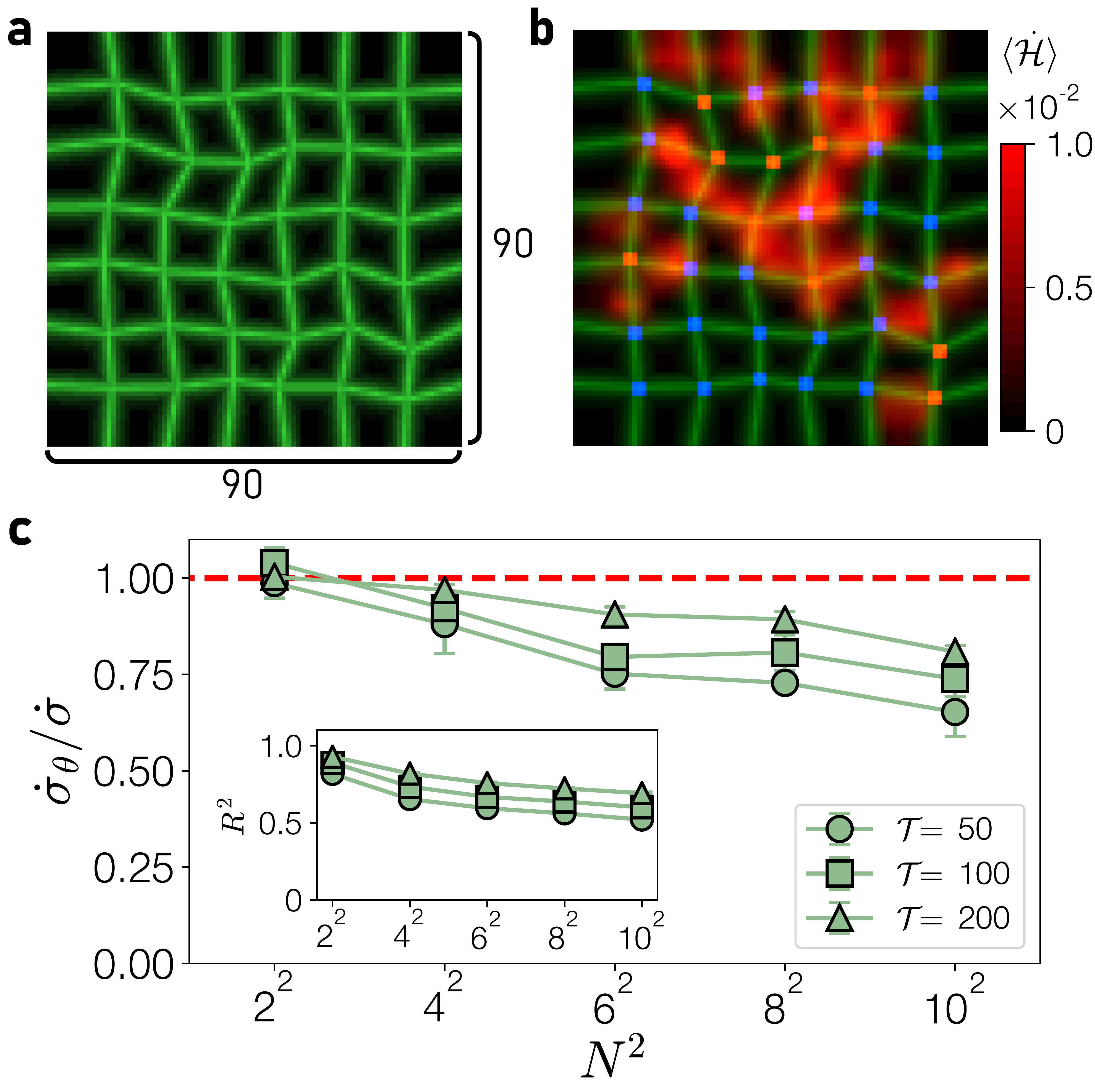}
    \vskip -0.1in
    \caption{{\bf a.} Snapshot of a $6^2$ filamentous network. Each image is an integer-valued array with $(90, 90)$ pixels. 
    {\bf b.} Dissipation map $\langle \dot{\mathcal{H}} \rangle$ (red) overlaid with the snapshot (green). 
    Red and blue squares represent nodes of temperature $T_h$ and $T_c$, respectively.
    {\bf c.} $\dot{\sigma}_\theta/\dot{\sigma}$ with increasing $N$ and varying recorded time $\mathcal{T}$. The inset shows $R^2$ between $\Delta S_\theta$ and $\Delta S$. Error bars represent the standard deviation of $\dot{\sigma}_\theta/\dot{\sigma}$ from five independent estimators.
    }\label{fig4}
\end{figure}

As similar to the previous example, the trajectories of $\bm{z}$ for training and validation sets are sampled from Langevin simulations using the Euler method and then transformed into $(15N, 15N)$ arrays (Fig.~\ref{fig4}a, Methods, and SMs 4--8).
To verify the performance with increasing dimensionality, the CNEEP is applied to measure the EP rate increasing $N$ from $2$ to $4$, $6$, $8$, and $10$ at $\mathcal{T}=200$.
Note that increasing the dimensionality causes an exponential growth of the volume of the available state space, called the curse of dimensionality; hence, the estimation of EP in a high-dimensional system is a challenging task, and even greater for time-series images.
Nevertheless, as can be seen in Fig.~\ref{fig4}c, we observe that although $\dot{\sigma}_\theta/\dot{\sigma}$ becomes less accurate with increasing $N^2$, $\dot{\sigma}_\theta$ is sufficiently close to $\dot{\sigma}$; for all $N^2$, $\dot{\sigma}_\theta/\dot{\sigma}$ is greater than $0.8$.
A linear regression between $\Delta S_\theta$ and $\Delta S$ is performed, and it is confirmed that $\Delta S_\theta$ and $\Delta S$ are well fitted with $R^2$ greater than $0.7$ (inset).
We also check the convergence of our method with increasing $\mathcal{T}$, with the inset showing the estimations for increasing $N$ become more accurate with $\mathcal{T}$.

High dimensionality typically leads to another arduous task, which is the identification of the dissipation map for a complex system. Most systems for which we hope to know the spatiotemporal pattern of EP are in such a situation, and so we examine whether the CNEEP can extract the dissipation map of the filamentous network.
Figure~\ref{fig4}b shows $\langle \dot{\mathcal{H}}(x, y) \rangle$ for the $N=6$ model.
Here, $\langle \dot{\mathcal{H}}(x, y) \rangle$ is intensively captured on the filaments between two connected nodes of different temperatures, whereas $\langle \dot{\mathcal{H}}(x, y) \rangle$ is almost zero between nodes of the same temperature.
This heterogeneous dissipation map is consistent with that the averaged heat current exists at the non-zero temperature difference [Eq.~\eqref{eq:XEP}], and thus it is evident that $\langle \dot{\mathcal{H}}(x, y) \rangle$ really reflects the dissipation map of a network of elastic filaments.
Based on this correct dissipation map, we expect that our approach can be used to reveal the inherent source of nonequilibrium activity, for instance, different stochasticities in other complex biological assemblies.
Dissipation maps for the other $N$ models are plotted in Supplementary Note 4, with the $\mathcal{H}(x, y)$ along the corresponding movies in SMs 4--8.

\section{Incomplete scenarios}
\label{sec:5}
\begin{figure*}[!t]
    \includegraphics[width=\linewidth]{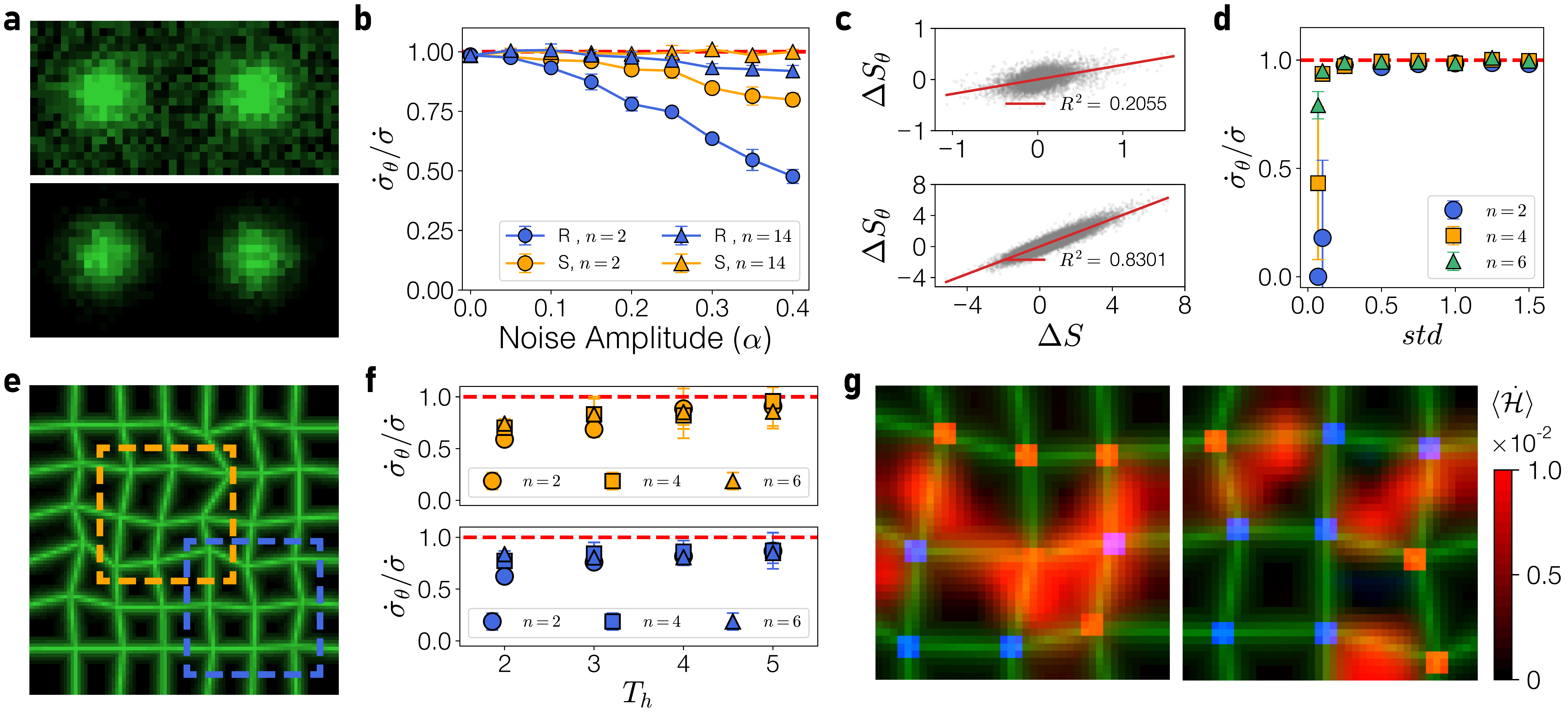}
    \vskip -0.1in
    \caption{
    {\bf a.} Noise-added images at a noise amplitude of $\alpha=0.4$ for an $N=2$ bead-spring model. The top and bottom panels represent a random and a shot noise case, respectively.
    {\bf b.} $\dot{\sigma}_\theta/\dot{\sigma}$ with increasing $\alpha$. The `R' (`S') symbol indicates the random (shot) noise case.
    {\bf c.} Scatter plots of $\Delta S$ and $\Delta S_\theta$ with their linear regression at random noise cases with $\alpha=0.4$ for $n=2$ (top) and $n=14$ (bottom).
    {\bf d.} $\dot{\sigma}_\theta/\dot{\sigma}$ with increasing standard deviation of bead displacements $std$ with varying $n$.
    {\bf e.} Snapshot of an $N=6$ filamentous network with orange and blue dashed squares, denoted `A1'  and `A2' in the text, indicating accessible areas.
    {\bf f.} $\dot{\sigma}_\theta/\dot{\sigma}$ as a function of $T_h$ with varying $n$ for the A1 (top) and A2 (bottom) areas.
    {\bf g.} Dissipation map $\langle \dot{\mathcal{H}} \rangle$ (red) for A1 (left) and A2 (right) overlaid with snapshots of the corresponding area (green). Red (blue) squares represent the nodes with temperature $T_h$ ($T_c$). In ({\bf b}), ({\bf d}), and ({\bf f}), error bars represent the standard deviation of $\dot{\sigma}_\theta/\dot{\sigma}$ from five independent estimators.
    }\label{fig5}
\end{figure*}

So far, we considered cases in which the images fully contained the information of the underlying system, but in real experiments, this is not often the case, where only incomplete information, e.g., noisy, coarse-grained, or partial observables from measurements, is available. To broaden our method's applicability to such incomplete scenarios, we alter the form of the input images to $\mathcal{I}_t^n$, which represents consecutive transitions with length $n$ defined by $\mathcal{I}_t^n \equiv \left( \mathcal{I}_t, \mathcal{I}_{t+1}, \dots, \mathcal{I}_{t+n-1} \right)$. That is, $\mathcal{I}_t^n$ is built by concatenating the images from $\mathcal{I}_t$ to $\mathcal{I}_{t+n-1}$. Following the definition of $\mathcal{I}_t^n$, the output of CNEEP is represented by
\begin{align} \label{eq:estimator_n}
    \Delta S_{\theta}( \mathcal{I}_t^n ) \equiv h_{\theta}(\mathcal{I}_t^n ) - h_{\theta} ( \tilde{\mathcal{I}}_{t}^n),
\end{align}
where $\tilde{\mathcal{I}}_t^n$ is the time-reversed trajectory of $\mathcal{I}_t^n$ and $h_\theta(\mathcal{I}_t^n )$ is the output of the CNN for the input $\mathcal{I}_t^n$.
By optimizing $J(\theta)$ using Eq.~\eqref{eq:lossftn}, $\Delta S_\theta (\mathcal{I}_t^n)$ converges to the coarse-grained EP $\Delta S (\mathcal{I}_t^n)$ along $\mathcal{I}_t^n$ given by~\cite{Roldan2010Estimating, kim2020learning}
\begin{align}
    \Delta S_\theta (\mathcal{I}_t^n ) = \ln{ \frac{ \mathcal{P}(\mathcal{I}_t^n )}{\mathcal{P}(\tilde{\mathcal{I}}_t^n )}}.
    \label{eq:sol_seq}
\end{align}
The right-hand-side is the definition of $\Delta S (\mathcal{I}_t^n)$.
When ${\mathcal{I}}_t$ is Markovian, $\Delta S (\mathcal{I}_t^n ) = \sum_{t'=t}^{t+n-2}\Delta S (\mathcal{I}_{t'}, \mathcal{I}_{t'+1})$. 
For non-Markovian systems such as a hidden Markov process, $\Delta S(\mathcal{I}_{t}, \mathcal{I}_{t+1})$ may vanish, and $\Delta S (\mathcal{I}_t^n )$ has been shown to be useful for obtaining the lower bound of the true EP~\cite{Roldan2010Estimating, martinez2019inferring, kim2020learning}.
According to this formalism, we get the estimations in this section by feeding $\mathcal{I}_t^n$ to the neural network.

To see the effects of noise on images, we consider two kinds of noise: random and shot noises. A random noise, which represents uncorrelated background noise, is uniformly sampled from a range $(0, 255\alpha)$, where $\alpha$ denotes the noise amplitude. 
By contrast, a shot noise is affected by the number of detected photons in optical devices, and thus it can be modeled by a Poisson process of detected photons in each pixel. Each shot noise is made to have the same variance as the corresponding random noise with amplitude $\alpha$, so the number of photons is in the range from $4800$ ($\alpha=0.05$) to $75$ ($\alpha=0.4$); refer to SMs 9, 10 for the noisy environments.

As illustrated in Fig.~\ref{fig5}a, the added noises blur the intensity of the pixels, preventing us from accessing the exact information of the underlying system, such as the bead displacements.
By these blurring effects, it becomes harder for the CNEEP with $n=2$ to estimate the EP rate with increasing $\alpha$, particularly for the random noise (Fig.~\ref{fig5}b).
Alternatively, we employ the CNEEP with $n=14$, and the results consistently show much higher performance; the $\dot{\sigma}_\theta/\dot{\sigma}$ for the random and shot noise are greater than $0.91$ and $0.99$ even at $\alpha=0.4$, respectively.
The scatter plots in Fig.~\ref{fig5}c also exhibit that $\Delta S_\theta$ with $n=14$ is better fitted to $\Delta S$ than $\Delta S_\theta$ with $n=2$.

Analogous to the role of noise, some relevant degree of freedom can be hidden or coarse-grained by an insufficient spatial resolution of an experimental microscope.
To cover this scenario, we generate movies of the bead-spring model with $N=2$ using normalized trajectories with varying standard deviation $std$ (SM 11 at $std=0.1$).
Since the movements of beads become more indistinguishable with decreasing $std$, we can control the spatial resolution of the movies by altering $std$ to the range $(0.07, 1.5)$ and examine the results with estimators of $n=2$, $4$, and $6$.
Remarkably, except for the CNEEP with $n=2$ at the extremely small $std \leq 0.1$, all of the estimators robustly infer the EP rate for the change of $std$ even at small $std$ (Fig.~\ref{fig5}d).

The partially observed scenario, in which only a partial area is accessible, is next implemented in a filamentous network with $N=6$, by extracting two fixed areas from the original images, `A1' and `A2', indicated as orange and blue dashed squares in Fig.~\ref{fig5}e. 
The partial area A1 (A2) is chosen to have a relatively larger (smaller) EP than the average.
With the assumption that one can only access a partial area of the whole system, we do not know how many other filaments exist or what the structures of the other parts look like.
With this condition, we employ the CNEEP with various $n$ for inferring the EP produced at the fixed areas, and markedly, the CNEEP successfully measures the partial EP for both areas (Fig.~\ref{fig5}f).
For more details, we test the CNEEP with increasing $T_h$ for varying $n$ and verify that $\dot{\sigma}_\theta/\dot{\sigma}$ decreases with decreasing $T_h$; namely, $\dot{\sigma}_\theta/\dot{\sigma} = 0.59$ and $0.62$ at $T_h=2$ for A1 and A2, respectively.
It may seem natural that the estimation is more difficult with smaller $\dot{\sigma}$, but we relieve this issue by using consecutive images with $n > 2$ by $\dot{\sigma}_\theta/\dot{\sigma} = 0.74$ and $0.84$ at $T_h=2$ for A1 and A2, respectively.
Figure~\ref{fig5}g reveals that $\langle \dot{\mathcal{H}}(x, y) \rangle$ at $T_h = 5$ for both A1 (left) and A2 (right) are distinctly concentrated at the filaments between the nodes of different temperatures as well as almost zero at other filaments that connect nodes of the same temperature.
These $\langle \dot{\mathcal{H}}(x, y) \rangle$ correspond to the results of the filamentous network in the previous section, thus demonstrating that the CNEEP is applicable to partially observed situations. $\mathcal{H}(x, y)$ along corresponding movies for A1 and A2 is shown in SM 12.

\section{Discussion}

Quantifying nonequilibrium activity and the extent to which energy is dissipated in diverse systems is essential to unveil the intrinsic system dynamics and energetics. While there have been many approaches to infer the dissipation, most studies have conventionally assumed that the relevant variables are known and tractable, so they cannot be applied to systems in which tracking is not possible or the relevant variables are unknown.
In this paper, we have introduced a new method called CNEEP to obtain a dissipation map as well as estimate the stochastic EP from time-series image data, without detailed information of the system or the process of taking out the trajectories of relevant variables. 

We have built the architecture of CNEEP, which allows us to directly reflect the spatial importance of EP, inspired by the visual interpretation techniques of CNNs~\cite{zhou2015learning, selvaraju2020gradcam}.
The CNEEP has been firstly applied to the bead-spring model to demonstrate its performance, and the results have shown that the CNEEP precisely captures how much EP occurs in different areas of images along a single movie as well as the ensemble average. 
After that, a filamentous network has been studied, where it was revealed that our method accurately captures the EP rate and dissipation map. This implies that the CNEEP can be practically employed in high-dimensional systems such as cytoskeletal networks and other multi-particle systems.

In addition, we have tested our method's applicability in cases where the measurement of the system's movements is hampered by noise, low spatial resolution, or partially observed areas, issues that frequently appear in real experiments. 
For more accurate estimation, consecutive images of length $n(>2)$ have been used as inputs, and as a result, we have confirmed that the CNEEP with $n(>2)$ is more robust to noise or spatial resolution than the estimator with $n=2$, and also has higher accuracy in low EP in partially observed areas.
Despite the improved performance from using consecutive transitions with length $n>2$, this method has a limitation in that the data required to cover the state space exponentially grows with increasing $n$.
Although the deep neural network leads us to effectively circumvent this issue, the method should be carefully utilized while monitoring the objective function $J(\theta)$ for the validation set (Supplementary Note 2).

Most notably, our work is the first study to the best of our knowledge to directly visualize dissipation from movies without preprocessing procedures.
The method can provide the particular areas that mainly produce the EP at each transition, and thus our study will offer a valuable tool for analyzing nonequilibrium system; for instance, distinguishing between active and thermal motion from recorded movies, finding the most important region for studying the energetics of the system, or identifying which interactions of particles cause dissipation.
We expect that our approach will be widely used as a tracking-free tool for inferring system dissipation only from movies in diverse fields, including complex living organisms, soft biological assemblies, or design efficient nano-devices.

\section{Methods}

We explain how we generate Brownian movies from models. 
Commonly, we collect $M$ trajectories with length $L$ of relevant model variables from overdamped Langevin simulations with a time interval of $\Delta \tau = 10^{-2}$. For the bead-spring and the filamentous network models, the collected trajectories are divided into two parts, a training set and a validation set. The training set is a directly used dataset for feeding the neural network, and the validation set is indirectly used for checking how well the model has been trained or for avoiding the overfitting problem (Supplementary Note 2). The ratio between training and validation sets is not strictly determined but dependent on users and the size of the dataset, so in this work we set the ratio to $4 : 1$.
After generating the trajectories, we perform the process of transforming the trajectories to movies consisting of $(W, H)$ arrays (i.e., images). 
Like general image data, each element of the array stands for the pixel intensity of the image, and all the elements have integer values from $0$ to $255$.
Details of the transforming process are written below for each model.

For the bead-spring model, the numerical simulation for sampling trajectories of the bead displacements $\bm{z}$ is governed by Eq.~\eqref{eq:Langevin} with fixed parameters $k = \gamma = 1$ and $T_h=10$. The temperature of the $i$-th bead $T_i$ is linearly varied from $T_c$ to $T_h$. These positional trajectories are normalized with a standard deviation $std$ and then transformed into $(20, 20 N)$ arrays, where $N$ is the number of beads in the model. We set $std=1$ by default, but in the last section, $std$ is set to vary from $0.07$ to $1.5$ to control the spatial resolution of the movies.
Notice that a displacement of $1$ of a trajectory corresponds to $1$ pixel in an image. Each displacement of a bead, $z_i$, is represented as a Gaussian function, which is generally used for modeling the point spread function of an imaging system~\cite{zhang2007gaussian, small2014fluorophore, manzo2015review}. Each Gaussian function takes a variance of $9$ and a maximum intensity of $255$, and is centered at $\left(10, 10 + 20 (i-1) + z_i \right)$. 

For a network of elastic filaments consisting of $N \times N$ nodes, each node moves in a two-dimensional plane and is connected to its nearest neighbors by elastic filaments. The trajectories of the node displacements are sampled from simulations with fixed parameters $k=\gamma=1$ and $T_c=1$ in Eq.~\eqref{eq:Langevin}.
Contrary to the bead-spring model, most nodes have temperature $T_c$, while others with temperature $T_h$ are randomly chosen with a probability $p=0.3$.
Figure~\ref{fig4} and corresponding results are described with $T_h=5$, but in the last section, $T_h$ varies from $1$ to $5$ to show the performance in a partially observed case.
In the transformation process, we draw the filaments between connected nodes making a grid in a $(15 N, 15 N)$ array.
The position of the nodes is determined by the sum of the grid position and their displacements $\bm{z}$.
The pixel intensities of the filaments decay with increasing distance, as a form of a Gaussian function with variance $1$ and maximum intensity $255$.

The noisy images in Fig.~\ref{fig5}a are constructed by adding a random or shot noise to the original images of the bead-spring model with $N=2$ and $T_c/T_h = 0.1$. The random noise is uniformly sampled from $0$ to $255\alpha$ regardless of the pixel location and intensity, where $\alpha$ is the noise amplitude in the range $[0, 1]$.
For the shot noise, which is dependent on the number of detected photons in each pixel,
the noisy pixel intensities are sampled from Poisson distributions with $\lambda(x,y)$, where the mean of the distribution $\lambda(x, y) = 12 \mathcal{I}(x, y)/\alpha^2 $ and $\mathcal{I}(x, y)$ is the pixel intensity at the spatial position $(x, y)$ on an image. 
Here, $12/\alpha^2$ is the maximum number of detected photons, determined to match the variance of the random noise with $\alpha$.
Lastly, we divide the image by $12/\alpha^2$ to be included in the range $[0, 255]$.
We saturate all the pixel intensities to $255$.
The partial images of the filamentous network with $N=6$ in Fig.~\ref{fig5}e are extracted from the original ones at the fixed areas A1 and A2. All the images are in the shape of $(40, 40)$ arrays.

\section{Data availability}
The datasets that support the result are available from the corresponding author upon request.

\section{Code availability}
The Python codes generating the dataset and for the CNEEP are available at \url{https://github.com/qodudrud/CNEEP}.

\begin{acknowledgments}
This study was supported by the Basic Science Research Program through the National Research Foundation of Korea (NRF Grant No. 2017R1A2B3006930).
\end{acknowledgments}

\section{Author contributions}
H.J. conceived the project. Y.B. and D.K.K. designed the research and wrote the codes. Y.B. performed the numerical simulations and the analytic calculations. All authors contributed to developing the network architecture, analyzing the results, and writing the paper.

\bibliography{main}



\section*{Supplementary material}

# Supplemental material: Attaining entropy production and dissipation maps from Brownian movies via neural networks

Youngkyoung Bae,[1] Dong-Kyum Kim,[1] and Hawoong Jeong[1, 2, *]

[1]*Department of Physics, Korea Advanced Institute of Science and Technology, Daejeon 34141, Korea*
[2]*Center for Complex Systems, Korea Advanced Institute of Science and Technology, Daejeon 34141, Korea*

## SUPPLEMENTARY NOTE 1: CALCULATION OF THE EXCHANGED ENTROPY PRODUCTION RATE

We derive the exchanged entropy production (EP) rate for the bead-spring model. The Langevin equation of the $N$ bead-spring model is given by

$$\gamma \dot{\boldsymbol{z}}(\tau) = \mathsf{A}\boldsymbol{z}(\tau) + \sqrt{2\mathsf{D}}\boldsymbol{\xi}(\tau), \tag{S1}$$

where $\boldsymbol{z}(\tau) = (z_1, z_2, \ldots, z_N)^T$ represents the displacements of $N$ beads at time $\tau$, and $\mathsf{A}$ and $\mathsf{D}$ are defined by $\mathsf{A}_{ij} = (\delta_{i,j+1} + \delta_{i,j-1} - 2\delta_{i,j})k$ with spring constant $k$ and $\mathsf{D} = \mathrm{diag}\{\gamma T_1, \gamma T_2, \ldots, \gamma T_N\}$ with $i$-th bead's temperature $T_i$, respectively. $\boldsymbol{\xi}$ is an independent Gaussian white noise vector satisfying $\langle \xi_i \rangle = 0$ and $\langle \xi_i(\tau)\xi_j(\tau') \rangle = \delta_{i,j}\delta(\tau - \tau')$. The angle bracket $\langle \cdot \rangle$ represents the ensemble average. From the fact that the system undergoes an Ornstein–Uhlenbeck process, the steady-state probability of $\boldsymbol{z}$ having a Gaussian form is $p(\boldsymbol{z}) \propto \exp\left[-1/2 \left(\boldsymbol{z}^T \mathsf{C}^{-1} \boldsymbol{z}\right)\right]$, where $\mathsf{C} \equiv \langle \boldsymbol{z}\boldsymbol{z}^T \rangle$ is the covariance matrix of $\boldsymbol{z}$, which can be obtained by solving the Lyapunov equation $\mathsf{AC} + \mathsf{CA} = -2\mathsf{D}$.

From a framework of stochastic energetics [1], the heat absorbed from the surrounding environment along the $i$-th bead over time interval $d\tau$ is given by

$$đQ_i = \left(-\gamma \dot{z}_i(\tau) + \sqrt{2\gamma T_i}\xi_i(\tau)\right) \circ dz_i(\tau), \tag{S2}$$

where $dz_i(\tau)$ is the change of $z_i(\tau)$ over $d\tau$ and $\circ$ denotes the Stratonovich product. Applying the Langevin equation [Eq. (S1)] and the definition of $\mathsf{A}$, $đQ_i$ can be written by

$$đQ_i = d\left(kz_i^2 - \frac{1}{2}\Theta[i-2]kz_{i-1}z_i - \frac{1}{2}\Theta[N-i-1]kz_i z_{i+1}\right)$$
$$- \frac{1}{2}\Theta[i-2]k\left(z_{i-1}\dot{z}_i - z_i \dot{z}_{i-1}\right)d\tau - \frac{1}{2}\Theta[N-i-1]k\left(z_{i+1}\dot{z}_i - z_i \dot{z}_{i+1}\right)d\tau, \tag{S3}$$

where $\Theta(n)$ equals 0 (1) if $n < 0$ ($n \geq 0$). The first term on the right-hand-side (RHS) is the contribution of the potential energy of the system, and hence it cannot affect the steady-state average. But the second (third) term, which is induced by a coupling between $z_i$ and $z_{i-1}$ ($z_{i+1}$), represents the exchanged heat between $z_i$ and $z_{i-1}$ ($z_{i+1}$), denoted by $đQ_{i,i-1}$ ($đQ_{i+1,i}$), and actually determines the ensemble averaged value [2–4]. Therefore, the heat current along $z_i$ in the steady state is given by

$$\langle \dot{Q}_i(\tau) \rangle = \Theta[i-2]\langle \dot{Q}_{i-1,i}(\tau) \rangle - \Theta[N-i-1]\langle \dot{Q}_{i,i+1}(\tau) \rangle. \tag{S4}$$

Here, we use the relation $\langle \dot{Q}_{i,i+1} \rangle + \langle \dot{Q}_{i+1,i} \rangle = 0$.

The EP is defined by the sum of medium entropy production and system entropy production, but in the steady state, the ensemble average of system entropy production is zero so that the EP rate can be given by

$$\dot{\sigma} = -\sum_{i=1}^{N} \frac{\langle \dot{Q}_i \rangle}{T_i} = \sum_{i=1}^{N-1} \langle \dot{Q}_{i,i+1} \rangle \left(\frac{1}{T_i} - \frac{1}{T_{i+1}}\right). \tag{S5}$$

Because $\langle \dot{Q}_{i,i+1} \rangle$ is the exchanged heat current between $z_i$ and $z_{i+1}$, the expression in the summation on the RHS becomes the exchanged EP rate $\dot{\sigma}_{i,i+1}$, which indicates the EP between $z_i$ and $z_{i+1}$. By the definition of $đQ_{i,i+1}$, $\langle \dot{Q}_{i,i+1} \rangle$ is obtained by

$$\langle \dot{Q}_{i,i+1} \rangle = -\frac{k}{2}\langle z_i v_{i+1} - z_{i+1} v_i \rangle, \tag{S6}$$

---

* hjeong@kaist.edu

where the local velocity $\boldsymbol{v}(\boldsymbol{z})$ is calculated as $\boldsymbol{v}(\boldsymbol{z}) = (\mathsf{A} + \mathsf{D}\mathsf{C}^{-1})\boldsymbol{z}$ in the linear system. Here, using $\boldsymbol{v}(\boldsymbol{z})$, $\langle z_i v_{i+1}\rangle$ is given by

$$\langle z_i v_{i+1}\rangle = \left\langle \left[\boldsymbol{z}\left((\mathsf{A}+\mathsf{D}\mathsf{C}^{-1})\boldsymbol{z}\right)^T\right]_{i,i+1}\right\rangle = \left(\mathsf{C}\mathsf{A}^T + \mathsf{D}\right)_{i,i+1}. \tag{S7}$$

Since $\mathsf{D}$ is a diagonal matrix, $\langle \dot{Q}_{i,i+1}\rangle$ is thus obtained by

$$\langle \dot{Q}_{i,i+1}\rangle = \frac{k}{2}\left(\mathsf{A}\mathsf{C} - \mathsf{C}\mathsf{A}^T\right)_{i,i+1}. \tag{S8}$$

Here, $\langle z_i v_{i+1} - z_{i+1} v_i\rangle = \left(\mathsf{C}\mathsf{A}^T - \mathsf{A}\mathsf{C}\right)_{i,i+1}$, so this term corresponds to the area enclosing rate or the specific angular momentum [4–7].

The numerical calculations of the exchanged EP between beads $i$ and $i+1$ along a trajectory, denoted by $S_{i,i+1}(\mathcal{T})$ with the recorded time $\mathcal{T}$ and plotted in Fig. 3 of the main text, are performed by

$$S_{i,i+1}(\mathcal{T}) = \frac{k}{2}\left(\frac{1}{T_i} - \frac{1}{T_{i+1}}\right)\sum_{t=0}^{L-1}\left(z_{i+1}^m(t\Delta\tau)z_i(t\Delta\tau) - z_i^m(t\Delta\tau)z_{i+1}(t\Delta\tau)\right), \tag{S9}$$

where $L$ is the number of steps, $\Delta\tau$ is a time interval between steps, and $z_i^m(t\Delta\tau) = (z_i(t\Delta\tau) + z_i((t+1)\Delta\tau))/2$.

## SUPPLEMENTARY NOTE 2: ALGORITHM

We follow the training procedure and algorithm written in our previous work [8]. Let $\mathcal{D}_{\text{train}}$ ($\mathcal{D}_{\text{val}}$) be a training (validation) dataset, consisting of images constructed from $M_{\text{train}}$ ($M_{\text{val}}$) trajectories with length $L$. Thus, $\mathcal{D}_{\text{train}}$ can be represented by $\mathcal{D}_{\text{train}} \equiv \{(\mathcal{I}_{m,0}, \mathcal{I}_{m,1}, \ldots, \mathcal{I}_{m,L-1})\}_{m=1,2,\ldots,M_{\text{train}}}$, where $\mathcal{I}_{m,t}$ denotes the image at step $t$ of the $m$-th trajectory. To train the CNEEP, we use the objective function $J(\theta)$ given by

$$J(\theta) = \langle \Delta S_\theta - e^{-\Delta S_\theta}\rangle, \tag{S10}$$

where $\Delta S_\theta$ is the output of our estimator, named the convolutional neural estimator for entropy production (CNEEP), and $\theta$ is the trainable parameters of the CNEEP. For instance, if $(\mathcal{I}_{m,t}, \mathcal{I}_{m,t+1})$ is fed to the CNEEP as an input, the output of CNEEP is denoted as $\Delta S_\theta(\mathcal{I}_{m,t}, \mathcal{I}_{m,t+1})$. To reduce computational cost, we uniformly sample a subset of $\mathcal{D}_{\text{train}}$, the so-called batch $\mathcal{B}$, and calculate $J_{\text{train}}(\theta)$ as Eq. (S11). After updating $\theta$ using the optimizer and the gradients of $J_{\text{train}}(\theta)$, we check the $J_{\text{val}}(\theta)$, evaluated from $\mathcal{D}_{\text{val}}$ as Eq. (S12). Here, we save the parameter $\theta^*$ by monitoring $J_{\text{val}}(\theta)$ to prevent the overfitting phenomenon. Figures S1 and S2 show the results with respect to training iteration for the bead-spring model and the filamentous network, respectively. As can be seen in Fig. S1, $\Delta S$ and $\Delta S_\theta$ are not aligned at all and $\dot\sigma_\theta$ is close to 0 before training, but as the training progresses, $\Delta S$ and $\Delta S_\theta$ are accurately fitted and $\dot\sigma_\theta$ quickly converges to the true rate $\dot\sigma$.

By contrast, for the filamentous network, the overfitting phenomenon appears due to high-dimensionality so that $J_{\text{train}}(\theta)$ constantly increases, whereas $J(\theta)$ decreases as training progresses. To resolve this overfitting problem, as we mentioned before, we save the parameter $\theta^*$ and equip CNEEP with it for measuring the EP or creating the dissipation map after training. Figure S2 shows the EP rate using the current parameter $\dot\sigma_\theta|_{\text{train}}$ and the best parameter $\dot\sigma_\theta|_{\text{best}}$, and as we expected, $\dot\sigma_\theta|_{\text{best}}$ well estimates $\dot\sigma$, whereas $\dot\sigma_\theta|_{\text{train}}$ diverges as training progresses.

## Algorithm 1 Training the CNEEP while monitoring $J_{\text{val}}(\theta)$

**Require:** The CNEEP $\Delta S_\theta$, optimizer, training set $\mathcal{D}_{\text{train}}$, validation set $\mathcal{D}_{\text{val}}$.
1: Initialize $J_{\text{best}} \leftarrow J_{\text{val}}(\theta)$, $\theta^\star \leftarrow \theta$.
2: **loop**
3:   Compute $J_{\text{train}}(\theta)$ from $\mathcal{D}_{\text{train}}$. $J_{\text{train}}(\theta)$ is calculated by

$$J_{\text{train}}(\theta) = \frac{1}{|\mathcal{B}|} \sum_{(m,t) \in \mathcal{B}} \left( \Delta S_\theta(\mathcal{I}_{m,t}, \mathcal{I}_{m,t+1}) - e^{-\Delta S_\theta(\mathcal{I}_{m,t}, \mathcal{I}_{m,t+1})} \right). \tag{S11}$$

4:   Compute gradients $\nabla_\theta J_{\text{train}}(\theta)$.
5:   Update parameters $\theta$ with the optimizer.
6:   Compute $J_{\text{val}}(\theta)$ from $\mathcal{D}_{\text{val}}$. $J_{\text{val}}(\theta)$ is calculated by

$$J_{\text{val}}(\theta) = \frac{1}{M_{\text{val}}(L-1)} \sum_{m=1}^{M_{\text{val}}} \sum_{t=0}^{L-1} \left( \Delta S_\theta(\mathcal{I}_{m,t}, \mathcal{I}_{m,t+1}) - e^{-\Delta S_\theta(\mathcal{I}_{m,t}, \mathcal{I}_{m,t+1})} \right). \tag{S12}$$

7:   **if** $J_{\text{val}}(\theta) > J_{\text{best}}$ **then**
8:     $J_{\text{best}} \leftarrow J_{\text{val}}(\theta)$
9:     $\theta^\star \leftarrow \theta$
10:   **end if**
11: **end loop**

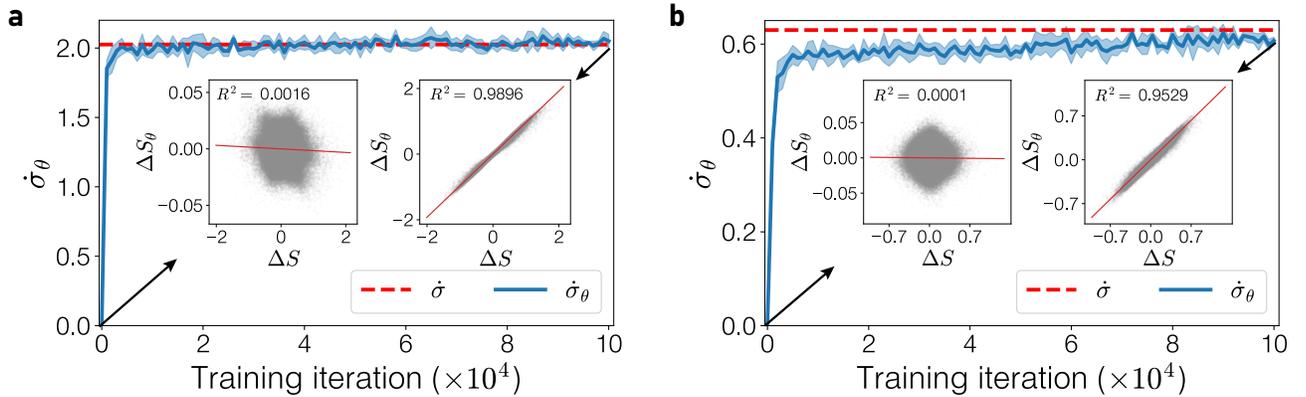

FIG. S1. **a, b.** $\dot{\sigma}_\theta$ with respect to training iteration for the bead-spring model with (**a**) $N=2$ and (**b**) $N=5$ at $T_c/T_h = 0.1$. The left (right) insets in both panels are scatter plots before (after) training between $\Delta S$ and $\Delta S_\theta$ with linear regression (solid red line) over all movies. The red dashed lines depict $\dot{\sigma}$. Shaded areas indicate the standard deviation of five independent estimators.

### SUPPLEMENTARY NOTE 3: TRAINING DETAILS OF THE CNEEP

The CNEEP consists of a convolutional neural network (CNN), the most popular class of neural networks for analyzing image data [9], to maximize the objective function $J(\theta)$ from Brownian movies. We use ReLU as the activation function and optimize $J(\theta)$ with the Adam optimizer [10]. Table I shows the details of the CNEEP architecture, where $\mathcal{I}_t^n \equiv (\mathcal{I}_t, \mathcal{I}_{t+1}, \ldots \mathcal{I}_{t+n-1})$ is the consecutive transitions with length $n$, $W$ and $H$ are the width and height of the images, respectively, and $C$ is the number of convolutional layer channels. Conv and Pool represent the convolutional layer and the max pooling layer. $(K, P, S)$ are the parameters of the convolutional layer, where $K$, $P$, and $S$ are the size of the filter, padding, and stride, respectively. In the input layer, the pixel values of input images $\mathcal{I}_t^n$ are normalized to have a zero mean and unit variance for training. In the Conv layers, $C$ is set to 32 and 128 for the bead-spring model and the filamentous network, respectively.

4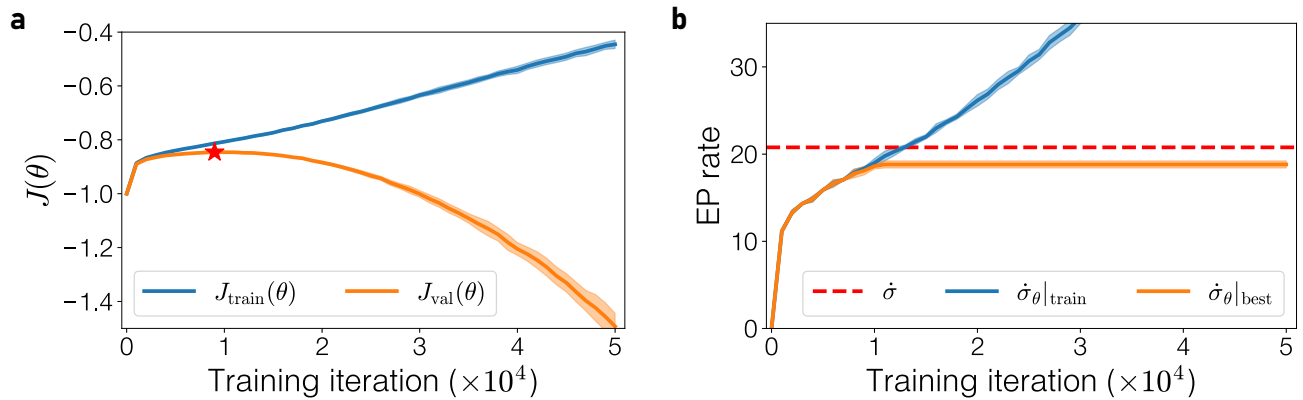

FIG. S2. **a.** $J(\theta)$ with respect to training iteration for the filamentous network with $N = 6$. The red star indicates the maximum value of $J_{\text{val}}(\theta)$ on average for five estimators. **b.** The EP rate estimation with respect to training iteration. $\dot{\sigma}_\theta|_{\text{train}}$ and $\dot{\sigma}_\theta|_{best}$ represent the EP rate estimations using the current parameters and the parameters at the maximum $J(\theta)$ [i.e. red star in (**a**)], respectively. The red dashed line depicts $\dot{\sigma}$. Shaded areas indicate the standard deviation of five independent estimators.

| CNEEP architecture | | | |
|---|---|---|---|
| Layer name | Output dim | $(K, P, S)$ | Act.Func. |
| Input $\mathcal{I}_t^n$ | $(n, W, H)$ | None | None |
| Conv + Pool | $(C, \lfloor W/4 \rfloor, \lfloor H/4 \rfloor)$ | $(5, 2, 2)$ | ReLU |
| Conv | $(C, \lfloor W/4 \rfloor, \lfloor H/4 \rfloor)$ | $(3, 1, 1)$ | ReLU |
| Conv | $(2C, \lfloor W/4 \rfloor, \lfloor H/4 \rfloor)$ | $(3, 1, 1)$ | ReLU |
| Output layer | 1 | None | None |

TABLE I. CNN configuration of the CNEEP: Layer name, output dimension of the layer, parameters of the convolutional layer $(K, P, S)$, and activation function.

## SUPPLEMENTARY NOTE 4: DISSIPATION MAP OF THE FILAMENTOUS NETWORK

We plot the dissipation map $\langle \dot{\mathcal{H}}(x,y) \rangle \equiv \langle \mathcal{H}(x,y)/\Delta\tau \rangle$ with $\Delta\tau = 10^{-2}$ for the network of elastic filaments with $N = 2, 4, 6, 8$, and $10$. Here, $\mathcal{H}(x, y)$ represents the spatial pattern of $\Delta S_\theta$, which is obtained by summing over the channel of the product of the last convolutional layer and the weight of the output layer [see Eqs. (4–5) in the main text]. As shown in all figures of Fig. S3, $\langle \dot{\mathcal{H}}(x,y) \rangle$ is mainly concentrated at the filaments connecting the nodes of different temperatures, while it vanishes at filaments connecting the nodes of the same temperature. These heterogeneous $\langle \dot{\mathcal{H}} \rangle$ are consistent with Eqs. (S5) and (S8), illustrating CNEEP's ability for capturing the real dissipation map. Please see the $\mathcal{H}(x, y)$ along a single movie for the bead-spring models and the filamentous networks in Supplementary Movies 3, 4, 6–10.

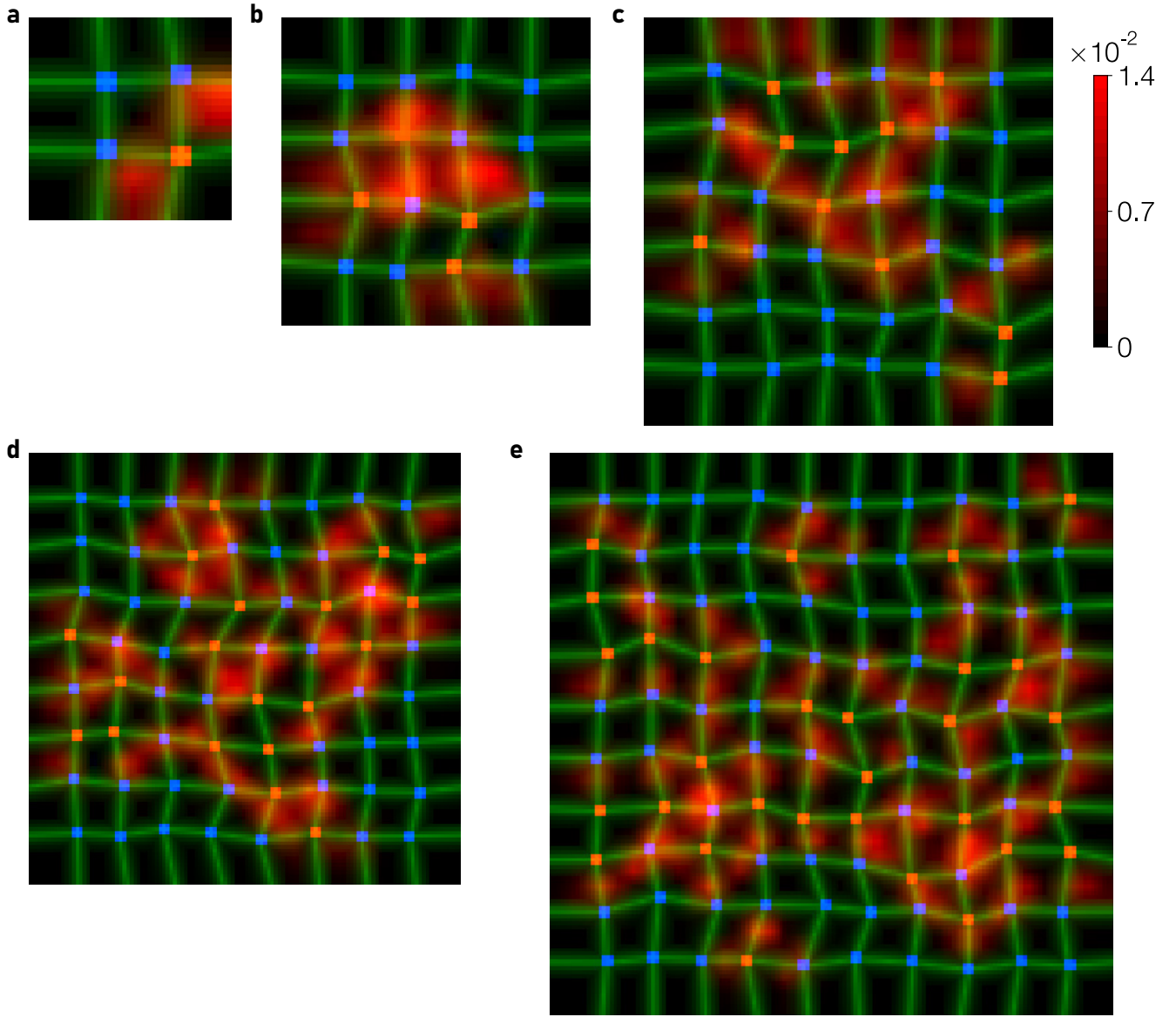

FIG. S3. Dissipation map $\langle \dot{\mathcal{H}}(x,y) \rangle$ of the filamentous networks with (**a**) $N = 2$, (**b**) 4, (**c**) 6, (**d**) 8, and (**e**) 10. $\langle \dot{H}(x,y) \rangle$ (red) overlaid with a snapshot of the filamentous network (green). Red and blue squares represent nodes of temperature $T_h$ and $T_c$, respectively. All figures share the same colorbar located at the top right.



\end{document}